\documentclass[11pt, reqno,titlepage]{article}
\usepackage{amsfonts}
\usepackage{amsmath,color}
\usepackage{ae,lscape}
\usepackage{hyperref}
\hypersetup{
    colorlinks=true,
    urlcolor=blue,
    linkcolor=blue,
    citecolor=blue
}

\usepackage[
backend=biber,
style=apa,
sorting=nyt
]{biblatex}
\DeclareSourcemap{
  \maps[datatype=bibtex]{
    \map{
      \step[fieldset=issn, null]
      \step[fieldset=doi, null]
      \step[fieldset=url, null]
      \step[fieldset=urldate, null]
    }
  }
}

\newtheorem{theorem}{Theorem}

\newtheorem{corollary}{Corollary}[section]

\newtheorem{example}{Example}

\newtheorem{lemma}{Lemma}[section]

\newtheorem{proposition}[theorem]{Proposition}
\newtheorem{remark}{Remark}[section]

\newtheorem{assumption}{Assumption}
\newenvironment{proof}[1][Proof]{\noindent\textbf{#1.} }{\ \rule{0.5em}{0.5em}}
\input{tcilatex}

\begin{document}

\title{\textbf{On Human Capital and Team Stability}\thanks{%
We thank Yeon-Koo Che, Navin Kartik, the editor and referees, and especially
Arnaud Dupuy, Fuhito Kojima and Phil Reny for useful comments. This paper
builds on material from an unpublished manuscript circulated under the name
``The Roommate Problem Is More Stable Than You Think,'' which is now
obsolete.

Accepted for publication by the \textit{Journal of Human Capital},
Volume 13, Number 2, Summer 2019. URL: \url{https://doi.org/10.1086/702925}.}}
\author{\textbf{Pierre-Andr\'{e} Chiappori}\thanks{%
Address: Department of Economics, Columbia University, 1009A International
Affairs Building, 420 West 118th St., New York, NY 10027, USA. E-mail:
pc2167@columbia.edu. Chiappori gratefully acknowledges financial support
from the NSF (award \#1124277), and ANR (grant Famineq).} \\
Columbia University \and \textbf{Alfred Galichon}\thanks{%
Economics Department and Courant Institute, New York University and
Economics Department, Sciences Po. Address: NYU, Department of Economics. 19
W 4th Street, New York, NY 10012, USA. Email: ag133@nyu.edu. Galichon
gratefully acknowledges funding from NSF grant DMS-1716489, ERC grants FP7-295298, 
FP7-312503, FP7-337665, and ANR grant Famineq.} \\
{NYU and Sciences Po} \and \textbf{Bernard Salani\'{e}}\thanks{%
Address: Department of Economics, Columbia University, 1131 International
Affairs Building, 420 West 118th Street, New York, NY 10027, USA. E-mail:
bsalanie@columbia.edu.} \\
Columbia University}
\date{August 13, 2017}
\maketitle

\begin{abstract}
In many economic contexts, agents from a same population team up to better
exploit their human capital. In such contexts (often called ``roommate
matching problems''), stable matchings may fail to exist even when utility
is transferable. We show that when each individual has a close substitute, a
stable matching can be implemented with minimal policy intervention. Our
results shed light on the stability of partnerships on the labor market.
Moreover, they imply that the tools crafted in empirical studies of the
marriage problem can easily be adapted to many roommate problems.
\end{abstract}

\section{Introduction}

Among Gary Becker's seminal contributions to labor economics, two are of
particular importance. The most obvious one is the notion of human capital,
after which this Journal is named. A second important contribution is the
development of matching models with transferable utility\footnote{%
See for instance \cite{Becker:73} and \cite{Becker:74}.}.
Although the market for marriage was Becker's favorite field of application
for the theory, many of the insights he developed are deeply relevant for
the analysis of labor issues as well. The labor market can often be
fruitfully seen as matching people to jobs---an insight that has  been
thoroughly exploited in the literature, in particular in a search context%
\footnote{%
See in particular \cite{MortensenPissarides94}.}.

There is, however, a fundamental difference between matching on the labor
and the marriage market. In the latter case, bilateral matching is a natural
framework; individuals who match mostly belong to two distinct
subpopulations. Not so, however, on the labor market. Workers match not only
to jobs, but also (and often primarily) to other workers. Lawyers gather in
law firms, doctors associate in medical practices, architects congregate in
architectural firms. While such partnerships are typical of the professions,
they extend to other services firms such as consultancies. More generally,
the notion that workers, through their employment relationships, tend to
match to other workers with similar characteristics, has received a lot of
attention and clear empirical support. For instance, a recent paper by %
\cite{EhrlichKim2015}  shows that immigrants endowed with similar
skills tend to team up and/or to separate in the same sectors. In
particular, higher levels of human capital of specific skill groups in the
destination country tend to increase the immigration flows of corresponding
groups from the source country. The authors convincingly argue that these
effects are crucial in assessing the economic consequences of migrations.

From a theoretical perspective, these features raise specific problems. In
sharp contrast with the bipartite literature, the formal analysis of
workers' matching on human capital must acknowledge the fact that the
individuals under consideration typically belong to the \emph{same
population.} It has been known for some time that this apparently minor
difference in settings may generate largely divergent properties. Take, for
instance, the specific case in which teams consist of exactly two people,
both coming from the same population: this is classically called the \emph{%
roommate matching problem}. 

The standard equilibrium concept in matching is
stability; a matching is stable if it is robust to unilateral and bilateral
deviations. We will follow this long tradition in this paper: when we say that ``an equilibrium exists'', for instance, we mean that ``a stable matching exists''.
Can we expect that the roommate matching game always has a
stable matching, so that the theoretical analysis could, as in the bipartite
framework, concentrate on the properties and the comparative statics of this
stable outcome? Or could it be the case that a stable matching fails to
exist, which might cast serious doubts on the relevance of matching models
for the analysis of these situations?

The answer to that important question has been known for a long time in the
Non Transferable Utility (NTU) context; indeed, \cite{Gale-Shapley:62}
have shown that stable matchings may not exist. However, applying the NTU
approach to a labor market requires wages to be exogenously fixed ,rather
than being endogenously determined at equilibrium. In most markets, this is
not the relevant framework.  Much more adequate is a Transferable Utility
(TU) framework, in which any potential team generates a surplus that is
(endogenously) shared by its members.

The problem, however, is that roommates matching games under TU tend to have
different properties than their bipartite counterparts. In a bipartite
setting, a stable matching exists under mild continuity and compactness
conditions; it maximizes aggregate surplus, and the associated individual
surpluses solve the dual imputation problem. A first conclusion of the
present paper is that in the two-partner roommate matching problem under TU,
on the contrary, stable matching may fail to exist. This is a potentially
damaging conclusion, since it might require reconsidering the relevance of
matching models in this context.

Our second conclusion, however, tends to mitigate this negative result by
showing that its economic implications may be much less damaging than one
would expect. Specifically, we consider a model in which agents belong to
various \textquotedblleft types\textquotedblright , where each type consists
of individuals of indistinguishable characteristics and tastes. In this
context, we show two main results. First, a stable matching always exists
when the number of individuals in each type is \emph{even}. Second, when the
number of individuals of any given type is large enough, there always exist
\textquotedblleft quasi-stable\textquotedblright\ matchings: even if a
stable matching does not exist, existence can be restored with minimal
policy intervention. To do this, one only needs to convince \emph{one}
individual to leave the game in each type with an odd number of individuals.
If this requires a compensation to be paid, this can be done at a per capita
cost that goes to zero when the population of each type goes to infinity.

We refer the reader to our conclusion for the implications of these findings
in terms of the stability of partnerships. We also show there that  the
empirical tools devised for the bipartite matching setting\footnote{%
See \cite{CSJEL:16} for a recent survey.} should carry over directly
to the roommate context when the populations under consideration are large.
Some of the results of the present paper are applied in this direction in %
\cite{CGG:15}.

\paragraph{Existing literature}

Since \cite{Gale-Shapley:62}, a few papers have studied the property
of NTU stable roommate matchings when they do exist. %
\cite{GusfieldIrving:89} showed that the set of singles is the same in
all stable matchings; \cite{KlausKlijn:10} study whether any of them
can be \textquotedblleft fair\textquotedblright . Efficient algorithms have
also been available since \cite{Irving:85}. Necessary and sufficient
existence conditions under strict preferences have been found by %
\cite{tan:91} for complete stable matchings and by %
\cite{Sotomayor:05} for stable matchings. \cite{kschung2000}
shows that a condition he calls \textquotedblleft no odd
rings\textquotedblright\ is sufficient for stable matchings to exist under
weak preferences. \cite{rodrineto07} introduces ``symmetric
utilities'' and \cite{Gudmundsson:2013ct} uses ``weak cycles.''

The TU case has been less studied in the theoretical literature, in spite of
its relevance in empirical applications. \cite{kschung2000} shows that
when the division of surplus obeys an exogenous rule, odd rings are ruled
out and the roommate problem has a stable matching; but that is clearly not
an appealing assumption. \cite{Karlander-Eriksson:01} provide a
graph-theoretic characterization of stable outcomes when they exist; and %
\cite{KN:10} studies their properties. \cite{Talman-Yang:11}
give a characterization in terms of integer programming.

The results of this paper are also related to those of \cite{AWW:13},
who show the existence of a Walrasian equilibrium in an economy with
indivisible goods, a continuum of agents and quasilinear utility. Unlike
their main results, ours apply in markets with finite numbers of agents. Our
methods are also original. As is well-known, in bipartite problems all
feasible matchings that maximize social surplus are stable. This is not true
in roommate problems; but we show how any roommate problem can be
\textquotedblleft cloned\textquotedblright\ in order to construct an
associated bipartite problem. We then exploit this insight to prove
existence of stable matchings in roommate problems with even numbers of
agents within each type.

To the best of our knowledge, the connection between the unipartite and
bipartite problems stressed in this paper is new.

\section{A Simple Example}

We start by giving the intuition of our main results on an illustrative
example.

\subsection{Unstable Matchings}

It has been known since Gale and Shapley that a stable matching may not
exist for the roommate problem under non-transferable utility. As it turns
out, it is almost equally easy to construct an example of non-existence of a
stable matching with transferable utility. Here a \emph{matching} defines
who is matched to whom \emph{and} how the corresponding surplus is divided
between the partners. Stability requires that

\begin{itemize}
\item no partner would be better off by leaving the partnership

\item no group of individuals could break off their current match, rematch
together, and generate a higher joint surplus than the sum of their current
individual utilities.
\end{itemize}

Consider the following example, in which only two-member matches are
possible:

\begin{example}
The population has three individuals. Any unmatched individual has zero
utility. The joint surplus created by the matching of any two of them is
given by the off-diagonal terms of the matrix%
\begin{equation}
\Phi =%
\begin{pmatrix}
- & 6 & 8 \\
6 & - & 5 \\
8 & 5 & -%
\end{pmatrix}
\label{MatrixPhi}
\end{equation}%
so that individuals 1 and 2 create, if they match, a surplus of 6; 1 and 3
create a surplus of 8, etc.

Assume that there exists a stable matching. A matching in which all
individuals remain single is obviously not stable; any stable matching must
be such that one person remains single and the other two are matched
together. Let $\left( u_{x}\right) $ be the utility that individual of type $%
x=1,2,3$ gets out of this game; stability imposes $u_{x}+u_{y}\geq \Phi
_{xy} $ for all potential matches, with equality if $x$ and $y$ are actually
matched---and $u_{x}\geq 0$ with equality if $x$ is single. One can readily
check, however, that no set of numbers $\left( u_{1},u_{2},u_{3}\right) $
satisfying these relationships for all $x$ and $y$ exists: whichever the
married pair is, one of the matched partners would increase her utility by
matching with the single person. Indeed, if the matched pair is $\left\{
1,2\right\} $, then
\begin{equation*}
u_{1}+u_{2}=6,u_{3}=0,u_{2}\geq 0
\end{equation*}%
contradicts $u_{1}+u_{3}\geq 8$: agent~3, being single, is willing to give
up any amount smaller than 8 to be matched with 1, while the match between 1
and 2 cannot provide 1 with more than 6. Similarly, if the married pair is $%
\left\{ 2,3\right\} $, then
\begin{equation*}
u_{2}+u_{3}=5,u_{1}=0,u_{2}\geq 0,u_{3}\geq 0
\end{equation*}%
contradicts both $u_{1}+u_{2}\geq 6$ and $u_{1}+u_{3}\geq 8$ (so that 1 is
willing to give more than 5 and less than 6 to agent~2 to match with her,
and more than 5 and less than 8 to 3.) Finally, if the married pair is $%
\left\{ 1,3\right\} $, then
\begin{equation*}
u_{1}+u_{3}=8,u_{2}=0,u_{1}\geq 0,u_{3}\geq 0
\end{equation*}%
is incompatible with $u_{1}+u_{3}\geq 11$, which follows from combining $%
u_{1}+u_{2}\geq 6$ and $u_{2}+u_{3}\geq 5$ with $u_{2}=0$ (since agent~2 is
single 1 could match with her and capture almost 6, while 3 could match with
her and capture almost 5; these outside options are more attractive than
anything 1 and 3 can achieve together.) We conclude that no stable matching
exists. \label{ex:noexistTU}
\end{example}

Note that there is nothing pathological in Example~\ref{ex:noexistTU}. The
surpluses can easily be (locally) modified without changing the result.
Also, the conclusion does not require an odd number of agents; one can
readily introduce a fourth individual, who generates a small enough surplus
with any roommate, without changing the non-existence finding.

\subsection{Cloning}

\label{subsec:cloning}

However, there exists a simple modification that restores existence in
Example~\ref{ex:noexistTU}. Let us now \emph{duplicate} the economy by
``cloning'' each agent; technically, we now have three \emph{types } $%
x=1,2,3 $ of agents, with two (identical) individuals of each type. The
joint surplus created by a matching between two individuals of different
types $x\neq y$ is as in Example~\ref{ex:noexistTU}; but we now also need to
define the surplus generated by the matching of two clones (two individuals
of the same type.) Take it to be 2 for every type---more on this later. We
then have the matrix:
\begin{equation}
\Phi ^{\prime }=%
\begin{pmatrix}
2 & 6 & 8 \\
6 & 2 & 5 \\
8 & 5 & 2%
\end{pmatrix}
\label{MatrixPhi2}
\end{equation}

Consider the following matching $\mu ^{\ast }$: there is one match between a
type 1 and a type 2 individuals, one between type 1 and type 3, and one
between type 2 and type 3. Assume individuals share the surplus so that each
individual of type 1 gets $4.5$, each individual of type 2 gets $1.5$, and
each individual of type 3 gets $3.5$. This is clearly feasible; and it is
easy to verify that it is a stable matching.

Less obvious but still true is the fact (proved later on) that existence
would still obtain for \emph{any} values chosen for the diagonal of the
matrix, although the stable matching pattern that would emerge may be
different\footnote{%
For instance, if the diagonal elements are large enough, the stable matching
matches each individual with her clone.}. In other words, our cloning
operation always restores the existence of a stable match, irrespective of
the values of the joint surpluses created by matches between clones.

\subsection{Surplus Maximization}

Our main result is better understood when related to another, closely linked
problem: finding a feasible matching that maximizes total surplus. Total
surplus is simply the sum of the joint surpluses of every match (keeping to
a normalized utility of zero for singles). In the standard, bipartite
framework, the adjective \textquotedblleft feasible\textquotedblright\
refers to the fact that each individual can only be matched to one partner
or stay single. Roommate matching, however, introduces an additional
feasibility constraint. For any two types $x\neq y$, denote $\mu _{xy}$ the
number of matches between an individual of type $x$ and an individual of
type $y$; since a roommate matching for which $\mu _{xy}$ and $\mu _{yx}$
differ would clearly not be feasible, it must be the case that $\mu
_{xy}=\mu _{yx}$. This additional symmetry constraint is absent from the
bipartite model, where these two individuals would belong to two separate
subpopulations and the number of marriages between say, a college-educated
man and a woman who is a high-school graduate may well differ (and typically
does) from the number of marriages between a college-educated woman and a
man who is a high-school graduate.

This symmetry constraint is the source of the difficulty in finding stable
roommate matchings; and our cloning operation addresses it. To see this on
our Example~\ref{ex:noexistTU}, first go back to roommate matching with one
individual of each type $x=1,2,3$, and neglect the symmetry constraint.
Since there is only one individual of each type $x$, she cannot match with
herself: $\mu _{xx}\equiv 0$; and neglecting symmetry, the only other
feasibility constraints are
\begin{equation*}
\mbox{for every }x,\;\;\sum_{y\neq x}\mu _{xy}\leq 1
\end{equation*}%
and
\begin{equation*}
\mbox{for every }y,\;\;\sum_{x\neq y}\mu _{xy}\leq 1.
\end{equation*}%
The two matchings
\begin{equation*}
\mu ^{1}=%
\begin{pmatrix}
0 & 0 & 1 \\
1 & 0 & 0 \\
0 & 1 & 0%
\end{pmatrix}%
\text{ and }\mu ^{2}=%
\begin{pmatrix}
0 & 1 & 0 \\
0 & 0 & 1 \\
1 & 0 & 0%
\end{pmatrix}%
\end{equation*}%
are feasible in this limited sense; and they both achieve the highest
possible surplus when the symmetry conditions are disregarded. The existence
of two solutions is not surprising: given the symmetric nature of the
surplus matrix $\Phi $, if a matrix $\mu $ maximizes total surplus, so does
its transpose $\mu ^{t}$. Unfortunately, neither is symmetric, and therefore
neither makes any sense in the roommate problem. For instance, $\mu ^{1}$
has agent~1 matched both with agent~3 (in the first row) and with agent~2
(in the first column). Also, note that a third solution to this relaxed
problem is the unweighted mean of $\mu ^{1}$ and $\mu ^{2}$,
\begin{equation*}
\mu ^{m}=%
\begin{pmatrix}
0 & 1/2 & 1/2 \\
1/2 & 0 & 1/2 \\
1/2 & 1/2 & 0%
\end{pmatrix}%
\end{equation*}%
However, while this matrix is indeed symmetric, its coefficients are not
integer and thus it is not a feasible matching either; moreover, and quite
interestingly, it cannot be interpreted as the outcome of randomization
since it is \emph{not} a convex combination of feasible roommate matching
matrices\footnote{%
For any stable roommate matching matrix, the sum of coefficients equals 2,
reflecting the fact that one agent must remain single. This property is
preserved by convex combination; however, the sum of coefficients of $\mu
^{m}$ equals 3.}.

Let us now reintroduce the symmetry constraint. The (now fully) feasible
matching that maximizes total surplus can only have one matched pair and one
single; and the pair that should be matched clearly consists of individuals~$%
1$ and $3$:
\begin{equation*}
\bar{\mu}=%
\begin{pmatrix}
0 & 0 & 1 \\
0 & 0 & 0 \\
1 & 0 & 0%
\end{pmatrix}%
.
\end{equation*}%
Obviously, $\bar{\mu}$ is \emph{not} a solution to the maximization problem
without symmetry constraint; in other words, the symmetry constraint is
binding in this example.\ As we shall see below, this is characteristic of
situations in which the roommate matching problem with transferable utility
does not have a stable matching. Indeed, we prove in the next section that
\emph{a stable matching exists if and only if the symmetry constraint does
not bind}.

Now take the ``cloned'' version of Example~\ref{ex:noexistTU}, in which each
type $x$ has two individuals. It is easy to see that the solution to the
relaxed problem which neglects the symmetry constraint is the $\mu^{\ast}$
of section~\ref{subsec:cloning}, which is symmetric; therefore the symmetry
constraint does not bind, and a stable matching exists. This is a general
result: we shall see below that in any cloned roommate matching setup, at
least one solution to the relaxed problem is symmetric---which implies the
existence of a stable match.

\subsection{A Bipartite Interpretation}

The relaxed problem, in turn, has a natural interpretation in terms of
bipartite matching. Start from the three-agent Example~\ref{ex:noexistTU},
and define an \emph{associated bipartite matching problem\/} as follows:
clone the population again, but this time assign a label (such as
\textquotedblleft man\textquotedblright\ or \textquotedblleft
woman\textquotedblright ) to each of the two subpopulations. Then consider
the bipartite matching problem between these subpopulations of
\textquotedblleft men\textquotedblright\ and \textquotedblleft
women\textquotedblright , with the joint surplus matrix given by $\Phi
^{\prime }$ in (\ref{MatrixPhi2}).

By standard results, there always exists a stable matching in this
associated bipartite matching problem; and it maximizes the associated total
surplus. In our example, $\mu ^{1}$ and $\mu ^{2}$ are the two stable
matchings. Any convex combination such as $\mu ^{m}$ can be interpreted as a
randomization between these two matchings; it is natural to focus on $\mu
^{m}$ since it is the only symmetric one and feasible roommate matchings
must be symmetric. As remarked above, in the original roommate problem $\mu
^{m}$ cannot be stable, since it has non-integer element.

Now if the roommate matching problem is cloned we can proceed as in the
above paragraph, except that with twice the number of individuals we should
work with $2\mu ^{m}$. As an integer symmetric matrix, reinterpreted in the
cloned roommate matching setup, it defines a feasible roommate matching
which is stable---in fact it is the stable matching $\mu ^{\ast }$ of
section~\ref{subsec:cloning}. This construction is general: we shall see
below that any roommate matching problem in which the number of individuals
in each type is even has a symmetric stable match.

We now provide a formal derivation of these results.

\section{The Formal Setting\label{par:RoommateMtchg}}

\label{sec:formalsett}

We consider a population of individuals who belong to a finite set of types $%
\mathcal{X}$. Individuals of the same type are indistinguishable. We denote $%
n_{x}$ the number of individuals of type $x\in \mathcal{X}$, and
\begin{equation*}
N=\sum_{x\in \mathcal{X}}n_{x}
\end{equation*}%
the total size of the population.

Without loss of generality, we normalize the utilities of singles to be zero
throughout.

\subsection{Roommate Matching}

A match consists of two partners of types $x$ and $y$. An individual of any
type can be matched with any individual of the same or any other type, or
remain single. In particular, there is no restriction that matches only
involve two partners of different \textquotedblleft
genders.\textquotedblright

Let a match $\left\{ x,y\right\} $ generate a surplus $\Phi _{xy}$. In
principle the two partners could play different roles. In sections~\ref%
{sec:formalsett} and~\ref{sec:matchlarge} we will assume that they are in
fact symmetric within a match, so that $\Phi _{xy}$ is assumed to be a
symmetric function of $(x,y)$:

\begin{assumption}
\label{ass:phisym} The surplus $\Phi_{xy}$ is symmetric in $(x,y)$.
\end{assumption}

We show in section~\ref{sect:NoSymm} that, surprising as it may seem, there
is in fact no loss of generality in making this assumption. The intuition is
simple: if $\Phi _{xy}$ fails to be symmetric in $\left( x,y\right) $, so
that the partners' roles are not exchangeable, then they should choose their
roles so to maximize output. This boils down to replacing $\Phi_{xy}$ with
the symmetric $\max \left( \Phi _{xy},\Phi _{yx}\right)$. Thus our results
extend easily when we do not impose Assumption~\ref{ass:phisym}; but it is
easier to start from the symmetric case.

\bigskip

A matching can be described by a matrix of numbers $\left( \mu _{xy}\right) $
indexed by $x,y\in \mathcal{X}$, such that

\begin{itemize}
\item $\mu _{x0}$ is the number of singles of type $x$

\item when $y\neq 0$, $\mu _{xy}$ is the number of matches between types $x$
and $y$.
\end{itemize}

The numbers $\mu _{xy}$ should be integers; given Assumption~\ref{ass:phisym}%
, they should be symmetric in $(x,y)$; and they should satisfy the scarcity
constraints. More precisely, the number of individuals of type $x$ must
equal the number $\mu _{x0}$ of singles of type $x$, plus the number of
pairs in which only one partner has type $x$, plus twice the number of pairs
in which the two partners are of type $x$---since such a same-type pair has
two individuals of type $x$.

\bigskip

Finally, the set of \emph{feasible roommate matchings} is
\begin{equation}
\mathcal{P}\left( n\right) =\left\{ \mu =\left( \mu _{xy}\right) :%
\begin{pmatrix}
2\mu _{xx}+\sum_{y\neq x}\mu _{xy}\leq n_{x} \\
\mu _{xy}=\mu _{yx} \\
\mu _{xy}\in \mathbb{N}%
\end{pmatrix}%
\right\}  \label{PureMtchg}
\end{equation}

\subsection{TU stability and optimality}

We define an \emph{outcome} $\left( \mu ,u\right) $ as the specification of
a feasible roommate matching $\mu$ and an associated vector of payoffs $%
u_{x} $ to each individual of type $x$. These payoffs have to be feasible:
that is, the sum of payoffs across the population has to be equal to the
total output under the matching $\mu$. Now in a roommate matching $\mu$, the
total surplus created is\footnote{%
Note that in the second sum operator the pair $\left\{ x,y\right\} $ appears
twice, one time as $(x,y)$ and another time as $(y,x)$; but the joint
surplus $\Phi _{xy}$ it creates must only be counted once, hence the
division by 2.}

\begin{equation}
S_{R}(\mu ;\Phi )=\sum_{x}\mu _{xx}\Phi _{xx}+\sum_{x\neq y}\mu _{xy}\frac{%
\Phi _{xy}}{2}.  \label{socialWelfare}
\end{equation}%
This leads to the following definition of a feasible outcome: an outcome $%
\left( \mu ,u\right) $ is \emph{feasible} if $\mu $ is a feasible roommate
matching and
\begin{equation}
\sum_{x\in \mathcal{X}}n_{x}u_{x}=S_{R}(\mu ;\Phi ).  \label{feasible}
\end{equation}

We define stability as in \cite{Gale-Shapley:62}: an outcome $\left(
\mu ,u\right) $ is stable if it cannot be blocked by an individual or by a
pair of individuals. More precisely, an outcome $\left( \mu ,u\right) $ is
\emph{stable} if it is feasible, and if for any two types $x,y\in \mathcal{X}
$, (i)$~u_{x}\geq 0$, and (ii)~$u_{x}+u_{y}\geq \Phi _{xy}$. By extension, a
matching $\mu $ is called stable if there exists a payoff vector $\left(
u_{x}\right) $ such that the outcome $\left( \mu ,u\right) $ is stable.

In bipartite matching the problem of stability is equivalent to the problem
of \emph{optimality}: stable matchings maximize total surplus. Things are
obviously more complicated in roommate matchings---there always exist
surplus-maximizing matchings, but they may not be stable. The maximum of the
aggregate surplus over the set of feasible roommate matchings $\mathcal{P}%
(n) $ is
\begin{eqnarray}
\mathcal{W}_{\mathcal{P}}\left( n,\Phi \right) &=&\max S_R(\mu;\Phi)
\label{SWFPure} \\
s.t.~ &&2\mu _{xx}+\sum_{y\neq x}\mu _{xy}\leq n_{x}  \notag \\
&&\mu _{xy}=\mu _{yx}  \notag \\
&&\mu _{xy}\in \mathbb{N}.  \notag
\end{eqnarray}%
While no stable matching may actually achieve this value, it plays an
important role in our argument.

\subsection{The Associated Bipartite Matching Problem}

We shall now see that to every roommate matching problem we can associate a
bipartite matching problem which generates almost the same level of
aggregate surplus. More precisely, we will prove that for every vector of
populations of types $n=(n_{x})$ and every symmetric surplus function $\Phi
=(\Phi _{xy})$, the highest possible surplus in the roommate matching
problem is \textquotedblleft close to\textquotedblright\ that achieved in a
bipartite problem with mirror populations of men and women and half the
surplus function:
\begin{equation*}
\mathcal{W}_{\mathcal{P}}(n,\Phi )\simeq \mathcal{W}_{\mathcal{B}}\left(
n,n,\Phi /2\right) .
\end{equation*}%
where $\mathcal{W}_{\mathcal{B}}\left( n,n,\Phi/2 \right) $ is defined as
the maximal surplus of the bipartite matching problem:

\begin{eqnarray}
\mathcal{W}_{\mathcal{B}}\left( n,n,\Phi/2\right) &=&\max_{\displaystyle \nu
\in \mathcal{B}(n,n)}S_B(\nu;\Phi)  \label{EquivMP}
\end{eqnarray}
where $S_B(\nu;\Phi)=\sum_{x,y\in \mathcal{X}}\nu _{xy}\frac{\Phi _{xy}}{2}$
and $\mathcal{B}(n,n)$ is the set of feasible matchings in the bipartite
problem:
\begin{equation}
\mathcal{B}\left( n,n\right) = \left\{ \nu =\left( \nu _{xy}\right) :
\begin{pmatrix}
\sum_{y}\nu _{xy}\leq n_{x} \\
\sum_{x}\nu _{xy}\leq n_{y} \\
\nu _{xy}\in \mathbb{N}%
\end{pmatrix}
\right\}  \label{BipartMtchg}
\end{equation}

We also define stability for a feasible bipartite matching $\left( \nu
_{xy}\right) $ in the usual way: there must exist payoffs $\left(
u_{x},v_{y}\right) $ such that%
\begin{eqnarray}
S_B(\nu;\Phi) &=&\sum_{x\in \mathcal{X}}n_{x}u_{x}+\sum_{y\in \mathcal{X}%
}n_{y}v_{y}  \label{DualProg} \\
\ u_{x}+v_{y} &\geq &\frac{\Phi _{xy}}{2}  \notag \\
u_{x}\geq 0, &&v_{y}\geq 0  \notag
\end{eqnarray}

By classical results of \cite{SS:72}, there exist stable matchings $%
\nu $, and they coincide with the solutions of (\ref{EquivMP}). Moreover,
the associated payoffs $(u,v)$ solve the dual program; that is, they
minimize $\sum_{x\in \mathcal{X}}n_{x}u_{x}+\sum_{y\in \mathcal{X}%
}n_{y}v_{y} $ over the feasible set of program (\ref{DualProg}). Finally,
for any stable matching, $\mu _{xy}>0$ implies $u_{x}+v_{y}=\Phi _{xy}/2$,
and $\mu _{x0}>0$ implies $u_{x}=0.$

\begin{remark}
The marriage problem obviously is a particular case of the roommate problem:
if in a roommate matching problem $\Phi _{xy}=-\infty $ whenever $x$ and $y$
have the same gender, then any optimal or stable matching will be
heterosexual.
\end{remark}

\subsubsection{Links Between $\mathcal{W}_{\mathcal{P}}$ and $\mathcal{W}_{%
\mathcal{B}}$}

It is not hard to see that $\mathcal{W}_{\mathcal{P}}\left( n,\Phi \right)
\leq \mathcal{W}_{\mathcal{B}}\left( n,n,\Phi /2\right) .$ In fact, we can
bound the difference between these two values:

\begin{theorem}
Under Assumption~\ref{ass:phisym}, \label{thm:controlSWF}
\begin{equation*}
\mathcal{W}_{\mathcal{P}}\left( n,\Phi \right) \leq \mathcal{W}_{\mathcal{B}%
}\left( n,n,\Phi/2 \right) \leq \mathcal{W}_{\mathcal{P}}\left( n,\Phi
\right) +\left\vert \mathcal{X}\right\vert ^{2}\overline{\Phi }
\end{equation*}%
where%
\begin{equation*}
\overline{\Phi }=\sup_{x,y\in \mathcal{X}}\Phi _{xy}.
\end{equation*}%
and $\left\vert \mathcal{X}\right\vert $ is the cardinal of the set $%
\mathcal{X}$, i.e. the number of types in the population.

\begin{proof}
See appendix.
\end{proof}
\end{theorem}

\

In some cases, $\mathcal{W}_{\mathcal{P}}\left( n,\Phi \right) $ and $%
\mathcal{W}_{\mathcal{B}}\left( n,n,\Phi /2\right) $ actually coincide. For
instance:

\begin{proposition}
\label{prop:EvenTypes}If $n_{x}$ is even for each $x\in \mathcal{X}$, then
under Assumption~\ref{ass:phisym},
\begin{equation*}
\mathcal{W}_{\mathcal{P}}\left( n,\Phi \right) =\mathcal{W}_{\mathcal{B}%
}\left( n,n,\Phi /2\right) .
\end{equation*}

\begin{proof}
See appendix.
\end{proof}
\end{proposition}

\subsubsection{Stable Roommate Matchings}

The existence of stable roommate matchings is directly related to the
divergence of $\mathcal{W}_{\mathcal{P}}\left( n,\Phi \right) $ and $%
\mathcal{W}_{\mathcal{B}}\left( n,n,\Phi /2\right) $. Indeed, one has:

\begin{theorem}
\label{thm:StableMatchings} Under Assumption~\ref{ass:phisym},

(i) There exist stable roommate matchings if and only if%
\begin{equation*}
\mathcal{W}_{\mathcal{P}}\left( n,\Phi \right) =\mathcal{W}_{\mathcal{B}%
}\left( n,n,\Phi /2\right) .
\end{equation*}

(ii) Whenever they exist, stable roommate matchings achieve the maximal
aggregate surplus $\mathcal{W}_{\mathcal{P}}\left( n,\Phi \right) $ in (\ref%
{SWFPure}).

(iii) Whenever a stable roommate matching exists, individual utilities at
equilibrium $\left( u_{x}\right)$ solve the following, dual program:
\begin{eqnarray}
\min_{u,A} &&\sum_{x}u_{x}n_{x}  \label{SWFFractDual} \\
s.t.~ &&u_{x}\geq 0  \notag \\
&&u_{x}+u_{y}\geq \Phi _{xy}+A_{xy}  \notag \\
&&A_{xy}=-A_{yx}  \notag
\end{eqnarray}
\end{theorem}

\begin{proof}
See appendix.
\end{proof}

\bigskip

Note that while the characterization of the existence of a stable matching
in terms of equality between an integer program and a linear program is a
well-known problem in the literature on matching (see %
\cite{Talman-Yang:11} for the roommate problem), the link with a
bipartite matching problem is new.

\bigskip

Also note that in program (\ref{SWFFractDual}), the antisymmetric matrix $A$
has a natural interpretation: $A_{xy}$ is the Lagrange multiplier of the
symmetry constraints $\mu _{xy}=\mu _{yx}$ in the initial program~(\ref%
{SWFPure}). Our proof shows that if $\mu _{xy}>0$ in a stable roommate
matching, then the corresponding $A_{xy}$ must be non-positive; but since $%
\mu _{yx}=\mu _{xy}$ the multiplier $A_{yx}$ must also be non-positive, so
that both must be zero. The lack of existence of a stable roommate matching
is therefore intimately linked to a binding symmetry constraint.

Given Proposition~\ref{prop:EvenTypes}, Theorem~\ref{thm:StableMatchings}
has an immediate corollary: with an even number of individuals per type,
there must exist a stable roommate matching. Formally:

\begin{corollary}
\label{prop:EvenTypesStable} If $n_{x}$ is even for each $x\in \mathcal{X}$,
then under Assumption~\ref{ass:phisym}, there exists a stable roommate
matching.
\end{corollary}

In particular, for any roommate matching problem, its \textquotedblleft
cloned\textquotedblright\ version, in which each agent has been replaced
with a couple of clones, has a stable matching; and this holds irrespective
of the surplus generated by the matching of two identical individuals. Of
course, in general much less than full cloning is needed to restore
existence; we give this statement a precise meaning in the next paragraph.

\bigskip

Our next result shows that one can restore the existence of a stable
matching by removing at most one individual of each type from the
population; if these individuals have to be compensated for leaving the
game, this can be done at limited total cost:

\begin{theorem}[Approximate stability]
\label{thm:approxStabl} Under Assumption~\ref{ass:phisym}, in a population
of $N$ individuals, there exists a subpopulation of at least $N-\left\vert
\mathcal{X}\right\vert $ individuals among which there exist a stable
matching, where $\left\vert \mathcal{X}\right\vert $ is the number of types.
The \emph{total }cost for the regulator to compensate the individuals left
aside is bounded above by $\left\vert \mathcal{X}\right\vert \overline{\Phi }%
.$
\end{theorem}

\begin{proof}
See appendix.
\end{proof}

\section{Matching in Large Numbers}

\label{sec:matchlarge}

We now consider the case of a ``large'' game, in which there are ``many''
agents \emph{of each type}. Intuitively, even though an odd number of agents
in any type may result in non existence of a stable roommate matching, the
resulting game becomes ``close'' to one in which a stable matching exists.
We now flesh out this intuition by providing a formal analysis.

We start with a formal definition of a large game. For that purpose, we
consider a sequence of games with the same number of types and the same
surplus matrix, but with increasing populations in each type. If $n_{x}^{k}$
denotes the population of type $x$ in game $k$ and $N^{k}=\sum_{x}n_{x}^{k}$
is the total population of that game, then we consider situations in which,
when $k\rightarrow \infty $:
\begin{equation*}
N^{k}\rightarrow \infty \text{ and }n_{x}^{k}/N^{k}\longrightarrow f_{x}
\end{equation*}%
where $f_{x}$ are constant numbers.

As the population gets larger, aggregate surplus increases proportionally;
it is therefore natural to consider the \emph{average surplus}, computed by
dividing aggregate surplus by the size of the population. We also extend the
definition of $\mathcal{W}_{\mathcal{B}}$ in program (\ref{EquivMP}) to
non-integers in the obvious way so as to define the limit average bipartite
problem $\mathcal{W}_{\mathcal{B}}\left( f,f,\Phi/2 \right)$. Note that the
linearity of the program implies
\begin{equation*}
\mathcal{W}_{\mathcal{B}}(cn,cm,\Phi/2 )=c\mathcal{W}_{\mathcal{B}%
}(n,m,\Phi/2 )
\end{equation*}%
for any $c>0$.

\begin{proposition}
\label{prop:BipartiteEquivLimit}In the large population limit, under
Assumption~\ref{ass:phisym}, the average surplus in the roommate matching
problem converges to the limit average surplus in the related bipartite
matching problem. That is,%
\begin{equation*}
\lim_{k\rightarrow \infty }\frac{\mathcal{W}_{\mathcal{P}}\left( n^{k},\Phi
\right) }{N^{k}}=\lim_{N^{k}\rightarrow \infty }\frac{\mathcal{W}_{\mathcal{B%
}}\left( n^{k},n^{k},\Phi /2\right) }{N^{k}}=\mathcal{W}_{\mathcal{B}}\left(
f,f,\Phi /2\right) .
\end{equation*}%
\newline
\end{proposition}

\begin{proof}
See appendix.
\end{proof}

Our approximation results crucially rely on the number of types becoming
small relative to the total number of individuals. By definition, two
individuals of the same type are indistinguishable in our formulation, both
in their preferences and in the way potential partners evaluate them. This
may seem rather strong; however, a closer look at the proof of Theorem~\ref%
{prop:BipartiteEquivLimit} shows that our bound can easily be refined. In
particular, we conjecture that with a continuum of types, Theorem~\ref%
{prop:BipartiteEquivLimit} would hold exactly.

\bigskip

A related effect of the number of individuals becoming much larger than the
number of types is that the costs of the policy to restore stability in
Theorem \ref{thm:approxStabl} become negligible:

\begin{proposition}
\label{prop:LargeLimitStable}In the large population limit and under
Assumption~\ref{ass:phisym},

(i) one may remove a subpopulation of asymptotically negligible size in
order to restore the existence of stable matchings.

(ii) the average cost per individual of restoring the existence of stable
matchings tends to zero.
\end{proposition}

\begin{proof}
See appendix.
\end{proof}

\bigskip

In particular, in the case of a continuum of individuals (that is, when
there is a finite number of types and an infinite number of individuals of
each type), we recover the results of \cite{AWW:13} (hereafter, AWW).
To make the connection with this paper, the partner types in our setting
translates into goods in AWW's. The social welfare in our setting translates
into the utility $u$ of a single consumer in AWW. $u$ is such that $u\left(
C\right) =\Phi \left( \left\{ x,y\right\} \right) $ for $C=\left\{
x,y\right\} $, $u\left( \left\{ x\right\} \right) =0$, and $u=-\infty $
elsewhere (or very negative). Then it can be shown without difficulty that
the existence of a TU\ stable matching in our setting is equivalent to the
existence of a Walrasian equilibrium in the AWW\ setting. Thus existence and
TU\ stability in the case of a continuum of individuals follows from Theorem
and Proposition in AWW.

\section{The Nonexchangeable Roommate Problem\label{sect:NoSymm}%
\protect\footnote{%
We are grateful to Arnaud Dupuy for correcting a mistake in a preliminary
version of the paper.}}

We now investigate what happens when the surplus $\Phi _{xy}$ is not
necessarily symmetric. This will arise when the roles played by the partners
are not exchangeable. For instance, a pilot and a copilot on a commercial
airplane have dissymmetric roles, but may be both chosen from the same
population. Hence, in this section, we shall assume away Assumption \ref%
{ass:phisym}, and we refer to the \textquotedblleft nonexchangeable roommate
problem\textquotedblright; it contains the exchangeable problem as a special
case.

As it turns out, this can be very easily recast in the terms of an
equivalent symmetric roommate problem. Indeed if $\Phi _{xy}>\Phi _{yx}$,
then any match of an (ordered) 2-uple $\left( y,x\right) $ will be dominated
by a matching of a $\left( x,y\right) $ 2-uple, and the partners may switch
the roles they play and generate more surplus. Therefore, in any optimal (or
stable) solution there cannot be such a $\left( y,x\right) $ 2-uple. As a
consequence, the nonexchangeable roommate problem is equivalent to an
exchangeable problem where the surplus function is equal to the maximum
joint surplus $x$ and $y$ may generate together, that is%
\begin{equation*}
\Phi _{xy}^{\prime }=\max \left( \Phi _{xy},\Phi _{yx}\right) ;
\end{equation*}%
and since this is symmetric our previous results apply almost directly.
Denoting $\pi _{xy}$ the number of $\left( x,y\right) $ pairs (in that
order), one has%
\begin{eqnarray*}
\mu _{xy} &=&\pi _{xy}+\pi _{yx},~x\neq y \\
\mu _{xx} &=&\pi _{xx}
\end{eqnarray*}%
and obviously, $\pi _{xy}$ need not equal $\pi _{yx}$. The population count
equation is%
\begin{equation*}
n_{x}=\sum_{y\in \mathcal{X}}\left( \pi _{xy}+\pi _{yx}\right) ,~\forall
x\in \mathcal{X}
\end{equation*}%
and the social surplus from a matching $\pi $ is
\begin{equation*}
\sum_{x,y\in \mathcal{X}}\pi _{xy}\Phi _{xy}.
\end{equation*}%
so that the optimal surplus in the nonexchangeable problem is%
\begin{eqnarray*}
\mathcal{W}_{\mathcal{P}}^{\prime}\left( n,\Phi \right) &=&\max \sum_{x,y\in
\mathcal{X}}\pi _{xy}\Phi _{xy} \\
&&s.t.~n_{x}=\sum_{y\in \mathcal{X}}\left( \pi _{xy}+\pi _{yx}\right)
,~\forall x\in \mathcal{X}.
\end{eqnarray*}

\bigskip

The following result extends our previous analysis to the nonexchangeable
setting:

\begin{theorem}
\label{thm:OptWelfareNonExch}The nonexchangeable roommate matching problem
is solved by considering the surplus function%
\begin{equation*}
\Phi _{xy}^{\prime }=\max \left( \Phi _{xy},\Phi _{yx}\right)
\end{equation*}%
which satisfies Assumption \ref{ass:phisym}. Call \emph{optimized symmetric
problem} the problem with surplus $\Phi _{xy}^{^{\prime }}$ and population
count $n_{x}$. Then:

(i) the optimal surplus in the nonexchangeable roommate problem coincides
with the optimal surplus in the corresponding optimized symmetric problem,
namely%
\begin{equation*}
\mathcal{W}_{\mathcal{P}}^{\prime}\left( n,\Phi \right) =\mathcal{W}_{%
\mathcal{P}}\left( n,\Phi ^{\prime }\right)
\end{equation*}

(ii) the nonexchangeable roommate problem has a stable matching if and only
if the optimized symmetric problem has a stable matching.
\end{theorem}

\bigskip

Given Theorem~\ref{thm:OptWelfareNonExch}, all results in Sections~\ref%
{sec:formalsett} and~\ref{sec:matchlarge} hold in the general
(nonexchangeable) case. In particular:

\begin{itemize}
\item Theorem~\ref{thm:controlSWF} extends to the general case: the social
surplus in the roommate problem with asymmetric surplus $\Phi _{xy}$ is
approximated by a bipartite problem with surplus function $\Phi
_{xy}^{\prime }=\max \left( \Phi _{xy},\Phi _{yx}\right) /2$, or more
formally:
\begin{equation*}
\mathcal{W}_{\mathcal{P}}^{\prime}\left( n,\Phi \right) \leq \mathcal{W}_{%
\mathcal{B}}\left( n,n,\Phi ^{\prime }/2\right) \leq \mathcal{W}_{\mathcal{P}%
}^{\prime}\left( n,\Phi \right) +\left\vert \mathcal{X}\right\vert ^{2}%
\overline{\Phi },
\end{equation*}%
and as an extension of Proposition~\ref{prop:EvenTypes}, equality holds in
particular when the number of individuals in each types are all even.

\item Theorem~\ref{thm:StableMatchings} extends as well: there is a stable
matching in the roommate problem with asymmetric surplus $\Phi _{xy}$ if and
only if there is equality in the first equality above, that is:%
\begin{equation*}
\mathcal{W}_{\mathcal{P}}^{\prime}\left( n,\Phi \right) =\mathcal{W}_{%
\mathcal{B}}\left( n,n,\Phi ^{\prime }/2\right) .
\end{equation*}

\item All the asymptotic results in Section~\ref{sec:matchlarge} hold true:
in the asymmetric roommate problem, there is approximate stability and the
optimal matching solves a linear programming problem.
\end{itemize}

\section{Conclusion}

From a technical perspective, our results are open to various extensions.
First, the empirical tools developed in the bipartite setting, especially
for the analysis of the marriage markets (see \cite{Choo-Siow:06}, %
\cite{CSW:17}, \cite{Fox:identmatchinggames}, %
\cite{Cupid:subRES}, to cite only a few\footnote{%
\cite{Graham:hdbk} has a good discussion of this burgeoning literature.%
}) can be extended to other contexts where the bipartite constraint is
relaxed. These include law firms or doctor practices, but also team jobs
such as pilot/copilot (and more generally team sports), as well as
\textquotedblleft tickets\textquotedblright\ in US presidential elections,
marriage markets incorporating single-sex households, and many others.

To be more specific, assume that the joint surplus $\tilde{\Phi}_{ij}$
generated by a match between two individuals $i$ (of type $x$) and $j$ (of
type $y$) is \emph{separable\/} in the sense defined by \cite{CSW:17}:
\begin{equation*}
\tilde{\Phi}_{ij}=\Phi_{xy}+\varepsilon^i_y+\eta^j_x;
\end{equation*}
separability assumes that unobserved heterogeneity terms do not interact in
the formation of joint surplus.

If the partnership has symmetric roles, then it is easy to see that $\Phi
_{xy}$ must be symmetric in $(x,y)$, and that $\eta $ and $\varepsilon $
must be the same family of random variables:
\begin{equation*}
\tilde{\Phi}_{ij}=\Phi _{xy}+\varepsilon _{y}^{i}+\varepsilon _{x}^{j}.
\end{equation*}%
Apart from this specific restriction, if there is a stable matching then the
results of \cite{CSW:17} apply: there exist $U$ and $V$ such that $%
U_{xy}+V_{xy}\equiv \Phi _{xy}$ and in equilibrium, if $i$ of type $x$ and $j
$ of type $y$ match then $i$ obtains surplus which stems from the
maximization of $U_{xz}+\varepsilon _{z}^{i}$ with respect to $z$, including
zero for the singlehood option in the maximization. In addition, symmetry
implies that $U_{xy}=V_{yx}$. This boils down the matching equilibrium to a
series of simple discrete-choice problems. We refer the reader to %
\cite{GS:AEAPP} for a short description of separable models, and to %
\cite{Cupid:subRES} for a much more complete study of identification
and estimation.

Secondly, while our analysis has been conducted in the discrete case, it
would be interesting to extend our results to the case where there is an
infinite number of agents with a continuum of types. We conjecture that this
could be done, at some cost in terms of the mathematics required\footnote{%
The relevant tools here come from the theory of optimal transportation, see %
\cite{Villani:2003} and \cite{McCannGuillen}. For the precise
connection between matching models and optimal transportation theory, see %
\cite{Ekeland:10}, \cite{GOZ:99} and %
\cite{ChiapporiMcCannNesheim:10}. It is also worth mentioning the
recent contribution of \cite{Ghoussoub-Moameni:13}, which uses the
same type of mathematical structure for very different purposes.}. Thirdly,
we conjecture that the same \textquotedblleft cloning\textquotedblright\
technique could be applied to matches involving more than two partners---the
multipartite reference, in that case, being the \textquotedblleft matching
for teams\textquotedblright\ context studied by %
\cite{CarlierEkeland:10}. Moreover, it seems natural to apply this
technique\ when utility is not transferable. One may think of assigning
arbitrarily genders to both clones of each type, and considering a bipartite
stable matching between the two genders. Such a matching will be stable in
the roommate matching framework if the bipartite matching of the cloned
populations is symmetric. However, such a symmetric stable bipartite
matching of the cloned population may not exist. Therefore, the usefulness
of cloning to restore stability in the non-transferable utility version of
the roommate problem is an open question.

Finally, some roommate problems involve extensions to situations where more
than two partners can form a match; but the two-partner case is a good place
to start the analysis. Here, we have shown that when the population is large
enough with respect to the number of observable types, the structure of the
roommate problem is the same as the structure of the bipartite matching
problem. Most empirical applications of matching models under TU use a
framework as in this paper in order to understand, depending on the context,
how the sorting on a given matching market depends on age, education or
income, but also height, BMI, marital preferences, etc.\footnote{%
See for instance \cite{Choo-Siow:06}, \cite{ChiappOreffice:08} %
\cite{ChiappOreffCFatter} among many others.}. We leave all this for
future research.

On a more substantive front, our conclusions are somewhat mixed. While
existence issues may be serious in specific contexts, large markets with a
discrete distribution of skills (or human capital) tend to be largely immune
from these problems. Specifically, and to put things a bit loosely,
two factors make partnerships between workers belonging to the same population more
likely to be stable: (i) when individuals can be clustered into a small number of
basic categories (the latter being defined by either a given level of human
capital or a specific combination of skills), and (ii) when different workers belonging
to the same category are \textquotedblleft close
substitutes\textquotedblright\ to each other---in the sense that substituting
one for the other does not change much the joint
surplus created in any partnership. For instance, we expect that medical
practices formed by a largish number of doctors with similar specialties
should be rather stable.

However, our results also imply that, in specific cases, stability may be a
serious concern. That would be the case when matching involves individuals
who do not have close substitutes, for instance because they each have a
rare and specific skill (think of a doctor who is the only expert available
on one particular disease). In professional partnerships (or in academia!),
management skills may also be very unevenly distributed; and our analysis
suggests that partnerships that depend on rare leadership skills are more
susceptible to break up.\footnote{%
Consultancies are an intermediate case: while junior consultants may be
relatively interchangeable, leadership matters in finding clients and
conserving them.} This would also true of firms that rely on a very
charismatic individual for inspiration. The early (1969--84) trajectory of
Apple under Steve Jobs may be a case in point. Last but not least, sport
teams involving a small number of superstars should exhibit stability
issues, especially when several stars are associated within the same team.
We are not aware of any systematic, empirical analysis of these issues;
however, it is fair to say that casual empiricism seems to support these
predictions.

When partnerships are least likely to be stable, firm-specific capital is
likely to stabilize a partnership. Regulation may also play a useful role.
While we do not pursue this here, one can imagine cases when non-compete or
\textquotedblleft no poaching\textquotedblright\ clauses that make mobility
more costly could actually be welfare-improving, if the courts allow them.
\cite[p.
330]{BeckerTreatiseFam} already cited the ability of homosexual unions to
\textquotedblleft dissolve without judiciary proceedings, alimony, or child
support payments\textquotedblright\ as one reason why they are less stable
than heterosexual unions. This is an interesting topic for further research.



\printbibliography

\newpage

\appendix

\section{Appendix: Proofs}

Our proofs use an auxiliary object: the highest possible surplus for a \emph{%
fractional\/} roommate matching, namely
\begin{equation}
\mathcal{W}_{\mathcal{F}}\left( n,\Phi \right) =\max_{\mu \in \mathcal{F}%
\left( n\right) }\left( \sum_{x}\mu _{xx}\Phi _{xx}+\sum_{x\neq y}\mu _{xy}%
\frac{\Phi _{xy}}{2}\right) .  \label{SWFFract}
\end{equation}%
where $\mathcal{F}\left( n\right) $ is the set of \emph{fractional
(roommate) matchings}, which relaxes the integrality constraint on $\mu $:
\begin{equation}
\mathcal{F}\left( n\right) =\left\{ \left( \mu _{xy}\right) :%
\begin{pmatrix}
2\mu _{xx}+\sum_{y\neq x}\mu _{xy}\leq n_{x} \\
\mu _{xy}=\mu _{yx} \\
\mu _{xy}\geq 0%
\end{pmatrix}%
\right\}.  \label{FractMtchg}
\end{equation}

The program (\ref{SWFFract}) has no immediate economic interpretation since
fractional roommate matchings are infeasible in the real world; and while
obviously $\mathcal{W}_{\mathcal{P}}\left( n,\Phi \right) \leq \mathcal{W}_{%
\mathcal{F}}\left( n,\Phi \right) $, the inequality in general is strict. We
are going to show, however, that the difference between the two programs
vanishes when the population becomes large. Moreover, we will establish a
link between (\ref{SWFFract}) and the surplus at the optimum of the
associated bipartite matching problem.

We start by proving:

\begin{lemma}
\label{lemma:wisw}
\begin{equation}
\mathcal{W}_{\mathcal{F}}\left( n,\Phi\right) =\mathcal{W}_{\mathcal{B}%
}\left( n,n,\Phi/2\right).
\end{equation}
Moreover, problem (\ref{SWFFract}) has a half-integral solution.
\end{lemma}

\begin{proof}[Proof of Lemma \protect\ref{lemma:wisw}]
First consider some fractional roommate matching $\mu \in \mathcal{F}\left(
n\right) $, and define
\begin{eqnarray*}
\nu _{xy} &=&\mu _{xy}\text{ if }x\neq y \\
\nu _{xx} &=&2\mu _{xx}.
\end{eqnarray*}%
As a (possibly fractional) bipartite matching, clearly $\nu \in \mathcal{B}%
\left( n,n\right) $; and
\begin{equation*}
\sum_{x}\mu _{xx}\Phi _{xx}+\sum_{x\neq y}\mu _{xy}\frac{\Phi _{xy}}{2}=%
\frac{1}{2}\sum_{x,y\in \mathcal{X}}\nu _{xy}\Phi _{xy}.
\end{equation*}%
Now the right-hand side is the aggregate surplus achieved by $\nu $ in the
bipartite matching problem with margins $(n,n)$ and surplus function $\Phi
/2 $. It follows that
\begin{equation}
\mathcal{W}_{\mathcal{F}}\left( n,\Phi \right) \leq \mathcal{W}_{\mathcal{B}%
}\left( n,n,\Phi /2\right) .  \label{OneIneq}
\end{equation}

Conversely, let $\left( \nu _{xy}\right) $ maximize aggregate surplus over $%
\mathcal{B}(n,n)$ with surplus $\Phi /2$. By symmetry of $\Phi $, $\left(
\nu _{yx}\right) $ also is a maximizer; and since (\ref{EquivMP}) is a
linear program, $\nu _{xy}^{\prime }=\frac{\nu _{xy}+\nu _{yx}}{2}$ also
maximizes it. Define
\begin{eqnarray*}
\mu _{xy}^{\prime } &=&\nu _{xy}^{\prime }\text{ if }x\neq y \\
\mu _{xx}^{\prime } &=&\frac{\nu _{xx}}{2}.
\end{eqnarray*}%
Then
\begin{eqnarray*}
2\mu _{xx}^{\prime }+\sum_{y\neq x}\mu _{xy}^{\prime } &=&\nu _{xx}+\frac{1}{%
2}\sum_{y\neq x}(\nu _{xy}+\nu _{yx}) \\
&=&\frac{1}{2}(\nu _{xx}+\sum_{y\neq x}\nu _{xy}) \\
&+&\frac{1}{2}(\nu _{xx}+\sum_{y\neq x}\nu _{yx}).
\end{eqnarray*}%
Now $\nu _{xx}+\sum_{y\neq x}\nu _{xy}\leq n_{x}$ by the scarcity constraint
of ``men'' of type $x$, and $\nu _{xx}+\sum_{y\neq x}\nu _{yx}\leq n_{x}$ by
the scarcity constraint of ``women'' of type $x$. It follows that $\mu
^{\prime }\in \mathcal{F}\left( n\right) $, and
\begin{equation*}
\sum_{x}\mu _{xx}^{\prime }\Phi _{xx}+\sum_{x\neq y}\mu _{xy}^{\prime }\frac{%
\Phi _{xy}}{2}=\frac{1}{2}\sum_{x,y\in \mathcal{X}}\nu _{xy}\Phi _{xy}.
\end{equation*}%
Therefore the values of the two programs coincide.

Half-integrality follows from the Birkhoff-von Neumann theorem: there always
exists an integral solution $\nu $ of the associated bipartite matching
problem, and the construction of $\mu ^{\prime }$ makes it half-integral%
\footnote{%
The half-integrality of the solution of problem (\ref{SWFFract}) also
follows from a general theorem of \cite{Balinski:1970}; but the proof
presented here is self-contained.}.
\end{proof}

\bigskip

Given Lemma~\ref{lemma:wisw}, we can now prove Theorem~\ref{thm:controlSWF}.

\bigskip

\begin{proof}[Proof of Theorem \protect\ref{thm:controlSWF}]
The first inequality simply follows from the fact that $\mathcal{P}\left(
n\right) \subset \mathcal{F}\left( n\right) $. Let us now show the second
inequality. Lemma~\ref{lemma:wisw} proved that $\mathcal{W}_{\mathcal{F}%
}\left( n,\Phi \right) =\mathcal{W}_{\mathcal{B}}\left( n,n,\Phi /2\right) $%
. Let $\mu $ achieve the maximum in $\mathcal{W}_{\mathcal{F}}\left( n,\Phi
\right) $, so that
\begin{equation*}
\mathcal{W}_{\mathcal{F}}\left( n,\Phi \right) =\sum_{x}\mu _{xx}\Phi
_{xx}+\sum_{x\neq y}\mu _{xy}\frac{\Phi _{xy}}{2}.
\end{equation*}%
Let $\left\lfloor x\right\rfloor $ denote the floor rounding of $x$; by
definition, $x<\left\lfloor x\right\rfloor +1$, so that
\begin{equation*}
\mathcal{W}_{\mathcal{F}}\left( n,\Phi \right) <\sum_{x}\left\lfloor \mu
_{xx}\right\rfloor \Phi _{xx}+\sum_{x\neq y}\left\lfloor \mu
_{xy}\right\rfloor \frac{\Phi _{xy}}{2}+\sum_{x}\Phi _{xx}+\sum_{x\neq y}%
\frac{\Phi _{xy}}{2}.
\end{equation*}%
The right-hand side can also be rewritten as
\begin{equation*}
\sum_{x,y}\left\lfloor \mu _{xy}\right\rfloor \Phi _{xy}+\sum_{x,y}\Phi
_{xy}.
\end{equation*}%
But $\left\lfloor \mu \right\rfloor $ is in $\mathcal{B}(n,n)$, and is
integer by construction; therefore
\begin{equation*}
\sum_{x,y\in \mathcal{X}}\left\lfloor \mu _{xy}\right\rfloor \Phi _{xy}\leq
\mathcal{W}_{\mathcal{P}}\left( n,\Phi \right) .
\end{equation*}%
Finally,
\begin{equation*}
\sum_{x,y\in \mathcal{X}}\Phi _{xy}\leq \left\vert \mathcal{X}\right\vert
^{2}\overline{\Phi }
\end{equation*}%
so that
\begin{equation*}
\mathcal{W}_{\mathcal{F}}\left( n,\Phi \right) \leq \mathcal{W}_{\mathcal{P}%
}\left( n,\Phi \right) +\left\vert \mathcal{X}\right\vert ^{2}\overline{\Phi
}.
\end{equation*}
\end{proof}

\subsection{Proof of Proposition \protect\ref{prop:EvenTypes}}

\begin{proof}
Let $n_{x}^{\prime }=\frac{n_{x}}{2}$. By Lemma~\ref{lemma:wisw}, problem $%
\mathcal{W}_{\mathcal{F}}\left( n^{\prime },\Phi\right) $ has an
half-integral solution $\mu^{\prime}$; therefore problem $\mathcal{W}_{%
\mathcal{F}}\left( n,\Phi\right) $ has an integral solution $2\mu^{\prime}$,
which must also solve (\ref{EquivMP}). It follows that
\begin{equation*}
\mathcal{W}_{\mathcal{P}}\left( n,\Phi\right) =\mathcal{W}_{\mathcal{F}%
}\left( n,\Phi\right) .
\end{equation*}
\end{proof}

\subsection{Proof of Theorem \protect\ref{thm:StableMatchings}}

\begin{proof}
By Theorem \ref{lemma:wisw}, Problem (\ref{SWFFract}) coincides with a
bipartite matching problem between marginal $\left( n_{x}\right) $ and
itself. By well-known results on bipartite matching, there exist vectors $%
\left( v_{x}\right) $ and $\left( w_{y}\right) $ such that%
\begin{eqnarray*}
v_{x}\geq 0, &&~w_{y}\geq 0 \\
v_{x}+w_{y} &\geq &\Phi _{xy}
\end{eqnarray*}%
and the latter inequality is an equality when $\mu _{xy}>0$. Setting%
\begin{equation*}
u_{x}=\frac{v_{x}+w_{x}}{2}
\end{equation*}%
the symmetry of $\Phi $ implies
\begin{eqnarray*}
u_{x} &\geq &0 \\
u_{x}+u_{y} &\geq &\Phi _{xy}
\end{eqnarray*}%
and%
\begin{equation*}
\sum_{x\in \mathcal{X}}n_{x}u_{x}=\sum_{x\in \mathcal{X}}\mu _{xx}\Phi
_{xx}+\sum_{x\neq y}\mu _{xy}\frac{\Phi _{xy}}{2}
\end{equation*}%
so that the outcome $(\mu,u)$ is stable.

Conversely, assume that $\mu $ is a stable roommate matching. Then by
definition, there is a vector $\left( u_{x}\right) $ such that%
\begin{eqnarray*}
u_{x} &\geq &0 \\
u_{x}+u_{y} &\geq &\Phi _{xy}
\end{eqnarray*}%
and
\begin{equation*}
\sum_{x\in \mathcal{X}}n_{x}u_{x}=\sum_{x\in \mathcal{X}}\mu _{xx}\Phi
_{xx}+\sum_{x\neq y}\mu _{xy}\frac{\Phi _{xy}}{2}.
\end{equation*}%
Therefore $\left( u,A=0\right) $ are Lagrange multipliers for the linear
programming problem (\ref{SWFFract}), and $\mu $ is an optimal solution of (%
\ref{SWFFract}); finally, $\mu$ is integral since it is a feasible roommate
matching. QED.

(i), (ii) and (iii) follow, as there exist integral solutions of (\ref%
{SWFFract}) if and only if
\begin{equation*}
\mathcal{W}_{\mathcal{P}}\left( n,\Phi\right) =\mathcal{W}_{\mathcal{F}%
}\left( n,\Phi\right) ,
\end{equation*}%
and $\mathcal{W}_{\mathcal{F}}\left( n,\Phi\right) $ coincides with $%
\mathcal{W}_{\mathcal{B}}\left( n,n,\Phi/2\right) $ from Lemma~\ref%
{lemma:wisw}.
\end{proof}

\subsection{Proof of Theorem \protect\ref{thm:approxStabl}}

\begin{proof}
For each type $x$, remove one individual of type $x$ to the population if $%
n_{x}$ is odd. The resulting subpopulation differs from the previous one by
at most $\left\vert \mathcal{X}\right\vert $ individuals, and there is an
even number of individuals of each type; hence by Proposition~\ref%
{prop:EvenTypesStable} there exists a stable matching.

Each individual so picked can be compensated with his payoff $u_{x}$. Since $%
u_{x}\leq \overline{\Phi }$, the total cost of compensating at most one
individual of each type is bounded from above by $\left\vert \mathcal{X}%
\right\vert \overline{\Phi }$.
\end{proof}

\subsection{Proof of Proposition \protect\ref{prop:BipartiteEquivLimit}}

\begin{proof}
By Theorem \ref{thm:controlSWF}, in the large population limit
\begin{equation*}
\lim_{k\rightarrow \infty }\frac{\mathcal{W}_{\mathcal{P}}\left( n^{k},\Phi
\right) }{N^{k}}=\mathcal{W}_{\mathcal{F}}\left( f,\Phi \right)
\end{equation*}%
and Lemma~\ref{lemma:wisw} yields the conclusion.
\end{proof}

\subsection{Proof of Proposition \protect\ref{prop:LargeLimitStable}}

\begin{proof}
(i) The number of individuals to be removed is bounded from above by $%
\left\vert \mathcal{X}\right\vert $, hence its frequency tends to zero as $%
\left\vert \mathcal{X}\right\vert /N\rightarrow 0$. (ii)\ follows from the
fact that
\begin{equation*}
\frac{\mathcal{W}_{\mathcal{F}}\left( n,\Phi \right) -\mathcal{W}_{\mathcal{P%
}}\left( n,\Phi \right) }{N}\rightarrow 0.
\end{equation*}
\end{proof}

\subsection{Proof of Theorem \protect\ref{thm:OptWelfareNonExch}}

\begin{proof}
(i) Consider an optimal solution $\mu _{xy}$ to $\mathcal{W}_{\mathcal{P}%
}\left( n,\Phi ^{\prime }\right) $. For any pair $x\neq y$ such that $\Phi
_{xy}>\Phi _{yx}$, set $\pi _{xy}=\mu _{xy}$, and $\pi _{xy}=0$ if $\Phi
_{xy}<\Phi _{yx}$. If $\Phi _{xy}=\Phi _{yx}$, set $\pi _{xy}$ and $\pi
_{yx} $ arbitrarily nonnegative integers such that $\pi _{xy}+\pi _{yx}=\mu
_{xy}$; set $\pi _{xx}=\mu _{xx}$. Then $\pi $ is feasible for the optimized
symmetric problem, and one has%
\begin{equation*}
\sum_{x\in \mathcal{X}}\mu _{xx}\Phi _{xx}^{\prime }+\sum_{x\neq y}\mu _{xy}%
\frac{\Phi _{xy}^{\prime }}{2}=\sum_{x,y\in \mathcal{X}}\pi _{xy}\Phi _{xy}
\end{equation*}%
so that%
\begin{equation*}
\mathcal{W}_{\mathcal{P}}\left( n,\Phi ^{\prime }\right) \leq \mathcal{W}_{%
\mathcal{P}}^{\prime}\left( n,\Phi \right) .
\end{equation*}

Conversely, consider $\pi _{xy}$ an optimal solution to $\mathcal{W}_{%
\mathcal{P}}^{\prime}\left( n,\Phi \right) $. First observe that if $\Phi
_{xy}<\Phi _{yx}$ then $\pi _{xy}=0$; otherwise subtracting one from $\mu
_{xy}$ and adding one to $\pi _{yx}$ would lead to an improving feasible
solution, contradicting the optimality of $\pi $. Set%
\begin{eqnarray*}
\mu _{xy} &=&\pi _{xy}+\pi _{yx},~x\neq y \\
\mu _{xx} &=&\pi _{xx}
\end{eqnarray*}%
so that
\begin{equation*}
\sum_{x\in \mathcal{X}}\mu _{xx}\Phi _{xx}^{\prime }+\sum_{x\neq y}\mu _{xy}%
\frac{\Phi _{xy}^{\prime }}{2}=\sum_{x,y\in \mathcal{X}}\pi _{xy}\Phi _{xy}
\end{equation*}%
and hence%
\begin{equation*}
\mathcal{W}_{\mathcal{P}}^{\prime}\left( n,\Phi \right) \leq \mathcal{W}_{%
\mathcal{P}}\left( n,\Phi ^{\prime }\right) .
\end{equation*}

\bigskip

(ii) Assume there is a stable matching $\pi _{xy}$ in the nonexchangeable
roommate problem. Then if there is a matched pair $\left( x,y\right) $ in
that order, one cannot have $\Phi _{yx}>\Phi _{xy}$; otherwise the coalition
$\left( y,x\right) $ would be blocking. Hence one can define
\begin{eqnarray*}
\mu _{xy} &=&\pi _{xy}+\pi _{yx},~x\neq y \\
\mu _{xx} &=&\pi _{xx}
\end{eqnarray*}%
and the matching $\mu $ is stable in the optimized symmetric problem.
Conversely, assume that the matching $\mu$ is stable in the optimized
symmetric problem. Then it is not hard to see that, defining $\pi $ from $%
\mu $ as in the first part of (i) above, the matching $\pi $ is stable in
the nonexchangeable roommate problem.
\end{proof}

\end{document}